\documentclass{article}
\usepackage{here,graphicx,subfigure}
\begin{document}
\title{ 
Minimal $Z'$ models and the early LHC \footnote{Based on a talk given at the 2$^{nd}$ Young Researchers Workshop \textit{Physics Challenges in the LHC Era}, Frascati, May 10 and 13, 2010.}}
\author{
Ennio Salvioni        \\
{\small \em Dipartimento di Fisica, Universit\`a di Padova and INFN,} \\
{\small \em Sezione di Padova, Via Marzolo 8, I-35131 Padova, Italy} \\
}
\date{}
\maketitle
\baselineskip=11.6pt
\begin{abstract}
We consider a class of minimal extensions of the Standard Model with an extra massive neutral gauge boson $Z'$. They include both family-universal models, where the extra $U(1)$ is associated with $(B-L)$, and non-universal models where the $Z'$ is coupled to a non-trivial linear combination of $B$ and the lepton flavours. After giving an estimate of the range of parameters compatible with a Grand Unified Theory, we present the current experimental bounds, discussing the interplay between electroweak precision tests and direct searches at the Tevatron. Finally, we assess the discovery potential of the early LHC. \end{abstract}
\baselineskip=14pt
\section{Introduction}
Extra neutral gauge bosons, known in the literature as $Z'$, appear in many proposals for Beyond-the-Standard Model (BSM) physics; for a review, see for instance \cite{Langacker}. Here we focus on \textit{minimal} $Z'$, previously studied in \cite{Appelquist}, which stand out both for their simplicity, and because they could arise in several of the above mentioned BSM scenarios, such as, \textit{e.g.}, Grand Unified Theories (GUTs) and string compactifications.  
\section{Theory}
Following \cite{Salvioni}, we consider a minimal extension of the SM gauge group that includes an additional Abelian factor, labeled $U(1)_{X}$, commuting with $SU(3)_{c}\times SU(2)_{L}\times U(1)_{Y}$. The fermion content of the SM is augmented by one right-handed neutrino per family. We require anomaly cancellation, as this allows us to write a renormalizable Lagrangian. If family-universality is imposed, then the anomaly cancellation conditions yield a unique solution: $X = (B-L)$, where $B$ and $L$ are baryon and lepton number respectively \footnote{The most general solution to the anomaly cancellation conditions is $X=a\,Y+b\,(B-L)$, with $a,\,b$ arbitrary coefficients. However, the $Y$ component can be absorbed in the kinetic mixing in the class of models we consider.}. However, if the requirement of family-universality is relaxed, it can be shown that the following set of family-dependent charges satisfy the anomaly cancellation conditions: 
$X= \sum_{a=e,\,\mu,\,\tau} (\lambda_{a}/3)(B-3L_{a})\,$,
where $L_{a}$ are the lepton flavours, and $\lambda_{a}$ are arbitrary coefficients. We will consider a specific example of such \textit{non-universal} $Z'$ in the following.  

In the basis of mass eigenstates for vectors, and with canonical kinetic terms, the neutral current Lagrangian reads 
\begin{equation}
\mathcal{L}_{NC}=e J^{\mu}_{em}A_{\mu}+g_{Z}\left(Z_{\mu}J_{Z}^{\mu}+Z'_{\mu}J_{Z'}^{\mu}\right)\,,
\end{equation}
where $A_{\mu}$ is the photon field coupled to the electromagnetic current, while $(Z_{\mu},\,Z'_{\mu})$ are the massive states, which couple to the currents
$(J_Z^\mu,\,J_{Z'}^\mu)$ respectively, obtained from
$J_{Z^0}^\mu  =  \, \sum_f \left[ T_{3L}(f) - \sin^2 \theta_W \, Q(f) \right]  \, \overline{f} \gamma^\mu f \,$ and
$J_{Z^{\prime \, 0}}^\mu  =  \sum_f \left[ g_Y \, Y(f) + g_{X} \, X (f) \right] \, \overline{f} \gamma^\mu f \,$ 
via a rotation of the $Z-Z'$ mixing angle $\theta'$. The explicit expression of the latter reads 
$\tan \theta^\prime = - ({g}_{Y}/{g_{Z}}) \, M_{Z^0}^2/({M_{Z^\prime}^2 - M_{Z^0}^2})\,$.
Thus, under our minimal assumptions, only three parameters beyond the SM ones are sufficient to describe the $Z'$ phenomenology: the physical mass of the extra vector, $M_{Z'}$, and the two coupling constants $(g_{Y},g_{X})$. In the following discussion, we normalize these couplings to the SM $Z^{0}$ coupling, namely $\widetilde{g}_{Y,X}= g_{Y,X}/g_{Z}$. 
\begin{figure} 
\centering
\includegraphics[width=0.7\textwidth]{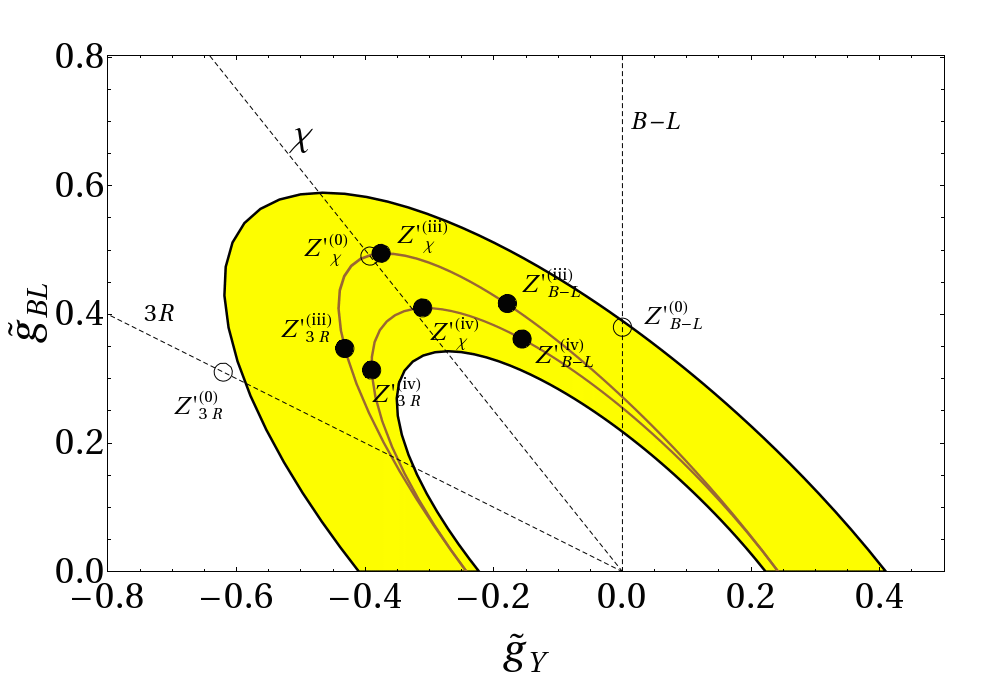}     
\caption{Estimate of the GUT-favoured region for the universal case, $X=(B-L)$. The yellow band represents the region of couplings compatible with a GUT, whereas dots and lines correspond to specific benchmark models or full supersymmetric GUT models, see \cite{Salvioni} for details.} 
\label{gut}
\end{figure}
\section{GUT-favoured region of parameters}
\label{sect:gut}
Because GUTs are one of the motivations for considering minimal $Z'$, it is interesting to give an estimate of the constraints that a GUT would imply on the weak-scale couplings $(\widetilde{g}_{Y},\widetilde{g}_{X})$. For choosing the boundary conditions at unification scale $M_{U}$, we normalize all charges as in $SO(10)$, and take $M_{U}=10^{16}$ GeV. We allow the $Z'$ coupling at unification scale to vary within the interval $1/100<g_{Z'}^{2}(M_{U})/(4\pi)<1/20$, and using the RGE of the model we obtain the GUT-favoured region of weak-scale couplings, shown in fig.\ref{gut} for the universal case $X=(B-L)$. Since the boundary conditions at scale $M_{U}$ are symmetric under the reflection $\widetilde{g}_{Y}\rightarrow -\widetilde{g}_{Y}$, it is evident from fig.~\ref{gut} that mixing effects in the RGE (due to the non-orthogonality of the generators $Y$ and $(B-L)$) are important. The GUT-favoured regions for non-universal models, computed along similar lines, can be found in \cite{Salvioni}. 
\section{Bounds from present data} \label{bounds}
The measurements providing constraints on minimal $Z'$ can be divided into two classes: electroweak precision tests and direct searches at the Tevatron. 
\subsection{Electroweak precision tests}
Measurements performed at LEP1 and at low energy mainly constrain $Z-Z'$ mixing, whereas data collected at LEP2 (above the $Z$ pole) constrain effective four-fermion operators. To compute the bounds from EWPT on minimal $Z'$, we integrate out the heavy vector and use the effective Lagrangian thus obtained to perform a global fit to the data. The results are shown in fig.~\ref{chi model}, for the universal `$\chi$ model', corresponding to a particular direction in the $(\widetilde{g}_{Y},\widetilde{g}_{X})$ plane often considered in the literature.
 
\subsection{Tevatron direct searches}
\begin{figure} 
\includegraphics[width=0.5\textwidth]{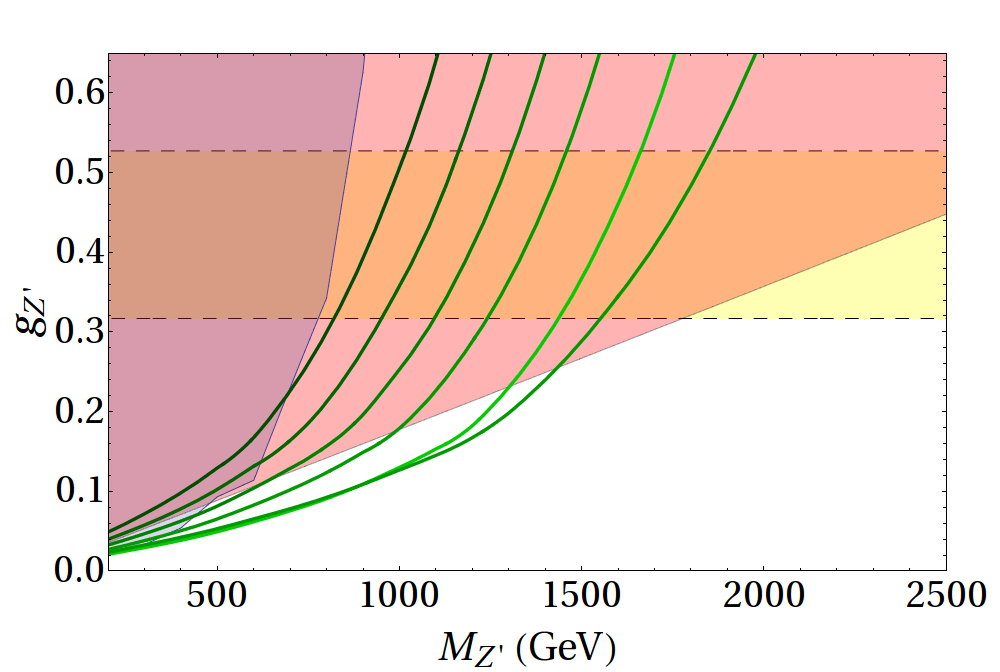}     
\includegraphics[width=0.5\textwidth]{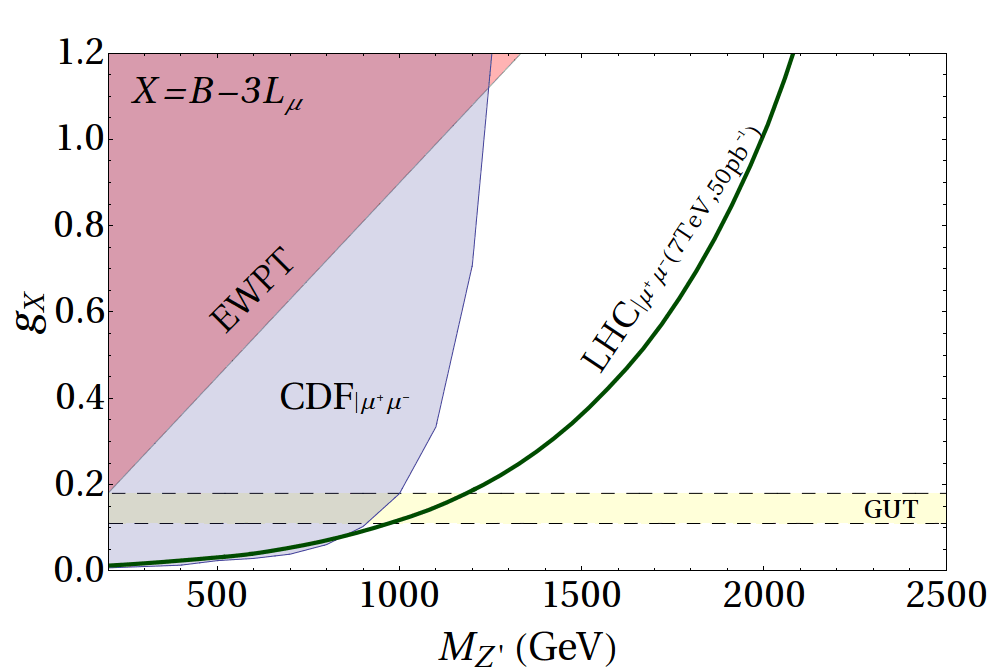}     

\caption{(left) Comparison of bounds from EWPT (red), Tevatron (blue), and discovery reach of the early LHC (green curves, from left to right: 50, 100, 200, 400 and 1000 pb${}^{-1}$ at 7 TeV, and 400 pb${}^{-1}$ at 10 TeV) for the $\chi$ model. (right) Present bounds and discovery prospects of the LHC at 7 TeV and 50 pb${}^{-1}$ for the \textit{muonphilic} model with $\widetilde{g}_{Y}=0$. For $g_{X}>0.3\,$, both the bounds from the Tevatron and the LHC reach are indeed weaker, because of finite-width effects not included in the figure, but the general message is unaffected. The yellow bands correspond to the GUT-favored region, see Section \ref{sect:gut}.}
\label{chi model}
\end{figure}

The CDF and D0 collaborations have derived, from the non-observation of discrepancies with the SM expectations, upper limits on $\sigma(\overline{p}p\rightarrow Z')\times Br(Z'\rightarrow \ell^{+}\ell^{-})$ ($\ell =e, \mu$), \cite{Aaltonen}. To extract bounds on minimal $Z'$, we compute the same quantity at NLO in QCD, and compare it with the limits published by the experimental collaborations. The comparison between bounds from EWPT and from the Tevatron is most clear if we plot them in (coupling \textit{vs.} mass), for a chosen direction in the $(\widetilde{g}_{Y},\widetilde{g}_{BL})$ plane, as it is done in fig.~\ref{chi model} for the $\chi$ model. We see that bounds from EWPT have a linear behaviour, because all the effects due to the $Z'$ in the low-energy effective Lagrangian depend on the ratio $g_{Z'}/M_{Z'}$, whereas bounds from the Tevatron become negligible above a kinematic limit, which is of the order of  1 TeV. Thus for low masses the Tevatron data give the strongest limits, while above a certain value of $M_{Z'}$ (which is of the order of 500 GeV for the $\chi$ model), bounds from EWPT are stronger. In particular, for models compatible with GUTs the strongest bounds are those given by EWPT.

\section{Early LHC reach}
The present schedule foresees that in 2010/2011 the LHC will run at 7 TeV in the center of mass, collecting up to 1 fb${}^{-1}$ of integrated luminosity $L$. Therefore, it is interesting to ask whether there are any minimal $Z'$ which are both allowed by present constraints and accessible for discovery in such early phase. To answer this question, we have performed a NLO analysis similar to the one used in extracting bounds from Tevatron data, requiring the $Z'$ signal to be at least a $5\sigma$ fluctuation over the SM-Drell Yan background. The results are displayed for the $\chi$ model in fig.~\ref{chi model}, where a comparison with present bounds is made. We see that for $L\sim 100$ pb${}^{-1}$ (the luminosity approximately foreseen at the end of 2010), no discovery is possible. On the other hand, for $L\sim1\,$fb${}^{-1}$ some unexplored regions become accessible; however, $Z'$ compatible with GUTs are still out of reach, and more energy and luminosity will be needed to test them. 

\subsection{The \textit{muonphilic} model}

We have seen that universal models are strongly constrained by present data. On the other hand, when we consider non-universal couplings to leptons, the bounds can be significantly altered. In particular, let us consider the case where $X=B-3L_{\mu}$, which we called `muonphilic $Z'\,$'. Let us further assume that kinetic mixing is negligible, \textit{i.e.} $\widetilde{g}_{Y}\approx 0$. In this case, the $Z'$ has no coupling to the first and third leptonic families, in particular it has no coupling to the electron. As a consequence, bounds from EWPT are strongly relaxed, the only surviving constraints coming from $(g-2)_{\mu}$ and $\nu$-$N$ scattering (NuTeV). On the other hand, the Tevatron reach is limited, as already noted in Section \ref{bounds}, to $M_{Z'}\leq 1$ TeV: therefore the LHC has access to a wide region of unexplored parameter space already with a very low integrated luminosity at 7 TeV, as shown in fig.~\ref{chi model}.

\section{Summary}

We have discussed the present experimental bounds and the early LHC reach on minimal $Z'$ models, showing that present constraints cannot be neglected when assessing the discovery potential of the early LHC. In particular, we have found that exploration of universal models, coupled to $(B-L)$, may need more energy and luminosity than those foreseen for 2010/2011, in particular for values of the couplings compatible with GUTs. On the other hand, some non-universal models which are weakly constrained by present data, such as the \textit{muonphilic} $Z'$, could be discovered at the LHC with very low integrated luminosity.
\section{Acknowledgements}
I am indebted to my collaborators F.~Zwirner, G.~Villadoro, and A.~Strumia. I would also like to thank the organizers of the 2$^{nd}$ Young Researchers Workshop \textit{Physics Challenges in the LHC Era} for giving me the possibility to present this work.

\end{document}